\documentclass[epsf]{kluwer}
\usepackage{epsfig}
\begin{document}
\begin{article}
\begin{opening}
\title{Numerical simulation of accretion discs in close binary systems and
discovery of spiral shocks}
\author{Takuya \surname{Matsuda} \email{tmatsuda@kobe-u.ac.jp}}
\author{Makoto \surname{Makita}}
\author{Hidekazu \surname{Fujiwara}}
\author{Takizo \surname{Nagae}} 
\author{Kei \surname{Haraguchi}}
\author{Eiji \surname{Hayashi}}
\institute{Department of Earth and Planetary Sciences, Kobe University, Kobe,
Japan}
\author{H.M.J. \surname{Boffin}}
\institute{Royal Observatory, Brussels, Belgium}

\begin{abstract}
The history of hydrodynamic numerical simulations for accretion disks in
close binary systems is reviewed, in which emphasis is placed, in particular,
on the facts that spiral shock waves were numerically found in 1986 by
researchers including one of the present authors and that spiral structure
was discovered in IP Pegasi in 1997 by Steeghs et al.  The results of our two
and three-dimensional numerical simulations in recent years are then
summarized, with comparison being made with observations.
\end{abstract}
\end{opening}

\section{Introduction}
The term ``accretion'' used in astrophysics means the infall of gas onto a
celestial body due to the gravitational attraction of the body.  On this
occasion, the gravitational energy of the gas is released and, eventually,
transformed into radiation and emitted to space.

The accretion process may be classified into two types depending on the momentum
possessed by the gas, i.e. whether the gas has a large angular momentum with
respect to the accreting object or not.  In the former case, a thin gaseous
disk, ``accretion disk'', is formed around the accreting object.  In such a
disk, the gas gradually loses its angular momentum by some mechanism and
accretes onto the central object.  On the other hand, a gas with less angular
momentum rapidly accretes onto the object directly and this is called ``wind
accretion'', a process characterized by, when the wind is supersonic,
the formation of ``bow shock'' in the forepart of the accreting object.  The
present paper deals with accretion disks.

Accretion disks are further classified into those in which the accreting body
has a mass comparable to that of typical star and those in which the
accreting body is a giant black hole expected to be present at the central
core of galaxy.  Our subject in this paper relates to the former.  With the
mass-accreting star being what is known as a compact star, which has a very
small size compared to its mass, such as black hole, neutron star or white
dwarf, a large amount of energy is released on accretion.  Such accretion
disks are present in close binaries, including cataclysmic variables, novae
and X-ray stars.

In accretion disks, for gas to accrete onto the central body, the gas must
lose, by some mechanism or another, its angular momentum.  The standard model
proposed by Shakura \& Sunyaev (1973) states that the angular momentum is
transported from the inner parts of the disk to the outer parts, due to some
kind of viscosity (see also Shakura 1972a, b, Pringle \& Rees 1972).  
The gas of the inner parts, having lost angular momentum, accretes onto 
the central star,
while a fraction of the outer gas, having obtained larger angular momentum,
transports it to infinity.  The gas can thus, almost entirely, accrete, while
conserving the total angular momentum.

A generally accepted candidate for this viscosity is turbulent viscosity. 
The standard model is often called the ``$\alpha$-disk model'', since the
magnitude of the turbulent viscosity has been characrerized by a
phenomenological parameter, $\alpha$.  For accretion disks, the Reynolds
number is as large as \(10^{11-14} \).  The usual hydrodynamical common sense
expects a flow with such a large Reynolds number to be turbulent.  With a
rotational fluid, however, the situation is not so simple.  Keplerian disks,
with angular momentum increasing outwards, amply satisfy the Rayleigh
criterion of stability.  That is, where the gas rotates with nearly Keplerian
motion, there is no convincing evidence so far for the accretion disk to
become unstable.  In contrast, there is available theoretical and numerical
evidence of the stability of Keplerian disks (Balbus, Hawley \& Stone, 1996).
Balbus \& Hawley (1991) concludes that a small magnetic field present in the
disk will be amplified during differential rotation and, eventually, generate
magnetic turbulence.  This has become one of the most popular models in recent
years.

Apart from the standard model, Sawada, Matsuda \& Hachisu (1986a, b, 1987)
conducted two-dimensional numerical simulations of accretion disk, to
discover the presence of spiral shock waves in the disk.  Spruit (1987) found
self-similar solutions having spiral shock waves.  They proposed a model in
which the spiral shock waves absorb angular momentum from the gas. This
is, eventually, transported to the orbital angular momentum of the binary
system, due to the axial asymmetry of the density distribution in the disk of
the gas and to the torque caused by the tidal force of the companion star. 
We call this the ``spiral shock model'', and we describe it in detail in the
present paper.
Many textbooks and reviews of accretion disks are available (for example,
Pringle 1981, Frank, King \& Raine 1992, Spruit 1995, Hartmann 1998,
Kato, Fukue \& Mineshige 1998).

\section{Historical overview of numerical simulation}
\subsection{two-dimensional hydrodynamic simulation of accretion disk}
The particle model was mainly used in early stages of numerical study for
accretion disks in close binaries.  Particles differ from fluid elements in
that the trajectories of the former can cross each other, while those of the
latter cannot.  Prendergast (1960) was the first to carry out numerical
simulation of gaseous flows, while ignoring the pressure.  His model
expressed the two constituent stars as two mass points and ignored either
release or accretion of the gas (see also Huang 1965, 1966).  These drawbacks
were corrected by Prendergast \& Taam (1974), who used the beam scheme and,
with the mass-accreting star having a large size, were unable to find
formation of any accretion disk.  Biermann (1971) conducted simulation by the
characteristic line method, for models which were close to  wind
accretion rather than accretion disks.

S\o rensen, Matsuda \& Sakurai (1974, 1975) made calculations using the Fluid in Cell
method (FLIC) and Cartesian coordinates.  They took both the mass-losing star
and the mass-accreting star into consideration and assumed the latter to be of
such a sufficiently small size as to allow formation of an accretion disk. 
The results of their calculation showed a gas stream from the L1 point
towards the compact star and formation of an accretion disk.  Flannery (1975)
performed a similar calculation and suggested the presence of a hot spot.

In contrast to the finite-difference method used by them, Lin \& Pringle
(1976) and Hensler (1982) performed calculations with a particle method, the former in
particular using the sticky particle method, which can be called a
predecessor of the SPH method.  All these calculations, having incorporated
an artificial viscosity to stabilize the calculation, could not reveal the
detailed structure of the inside of an accretion disk.

  Eleven years after the S\o rensen et al. (1975), Sawada, Matsuda \& Hachisu
(1986a, b, 1987) made the second challenge to the same problem, with use of
such state-of-the-art techniques as the Osher upwind finite-difference method
with 2nd order accuracy, generalized curvilinear coordinates and a super
computer of vector type.  The Osher upwind difference method can run the
calculation stably while suppressing the artificial viscosity at a low level
and is a predecessor of the TVD method, which is a representative modern
computational fluid dynamics scheme.  As a result, they discovered in the
accretion disk the presence of spiral shocks---the very feature having been
never discovered with use of other more dissipative schemes.

Since then, various authors have been carrying out two-dimensional simulations for
accretion disks by various methods and they all obtained spiral shocks
(Spruit et al. 1987, Rozyczka \& Spruit 1989, Matsuda et al. 1990, Savonije,
Papaloizou \& Lin 1994, Godon 1997).

Figure 1 shows the results of the two-dimensional simulation performed
recently by Makita, Miyawaki \& Matsuda (1998).  The mass ratio of the binary
is 1.  The region of calculation, which is limited to the surroundings of
the mass-accreting star, is in the range $[-0.5a, 0.5a] \times [-0.5a, 0.5a]$,
where $a$ denotes the separation of the binary.  The number of grid points is
$200 \times 200$.  Calculation is made for the specific heat ratio, $\gamma$, of
1.01, 1.05, 1.1 and 1.2, while using the equation of state for a perfect gas.
The lower $\gamma$ is used to take somehow cooling effects into account.

\begin{figure*}
\begin{center}
\begin{tabular*}{1.0\columnwidth}{p{0.5\columnwidth}p{0.5\columnwidth}}
\leavevmode
\epsfig{file=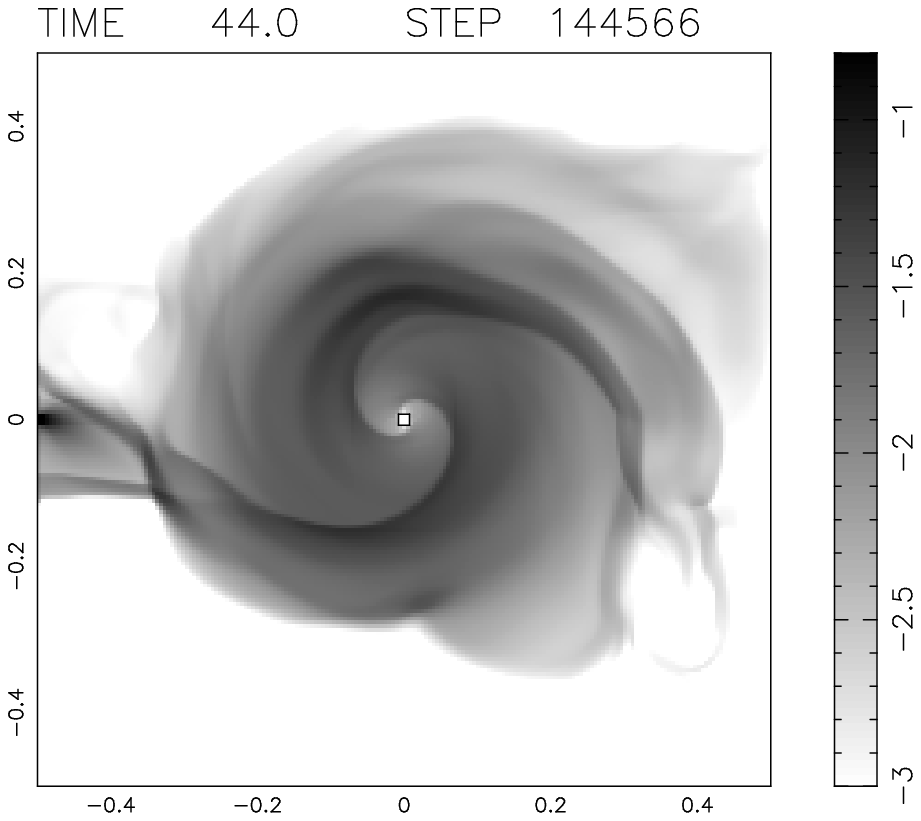, width=50mm}&
\hspace{-3mm}\epsfig{file=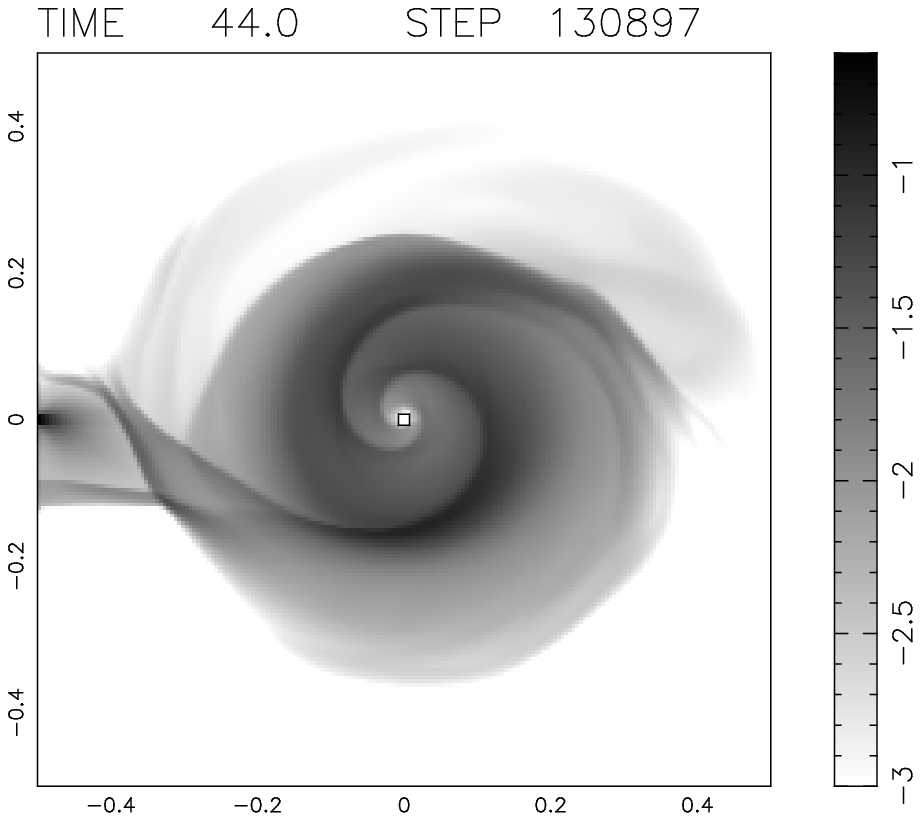, width=50mm}\\
\vspace{-5mm}\hspace{12mm}{\Large $\gamma=1.2$}&
\vspace{-5mm}\hspace{10mm}{\Large$\gamma=1.1$}\\
\epsfig{file=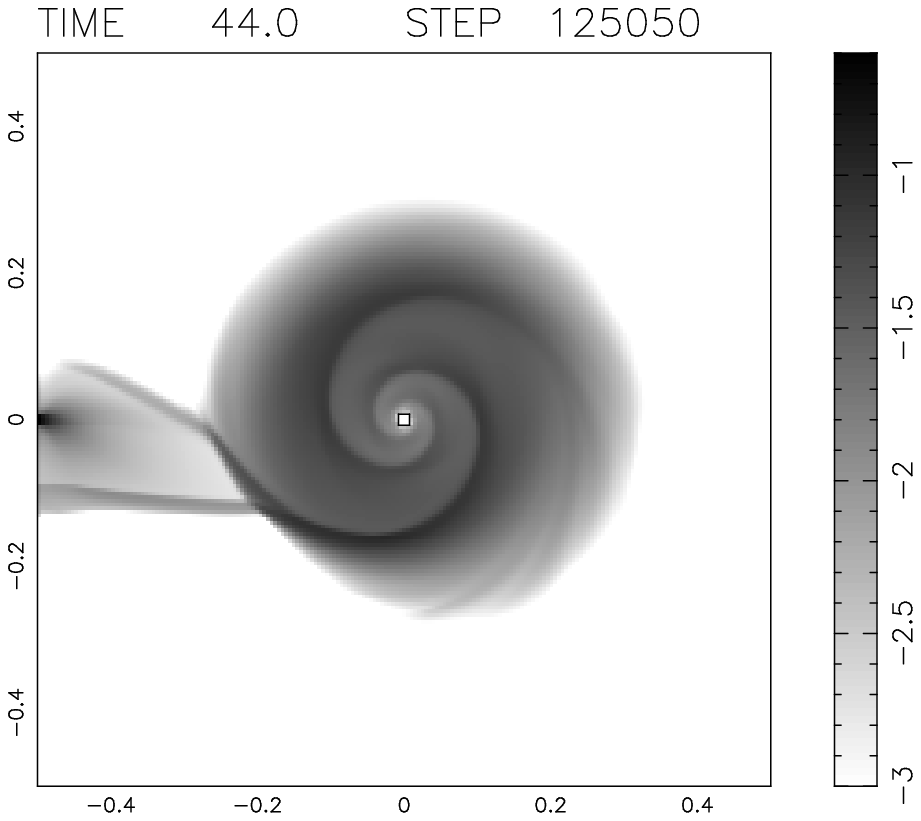, width=50mm}&
\hspace{-3mm}\epsfig{file=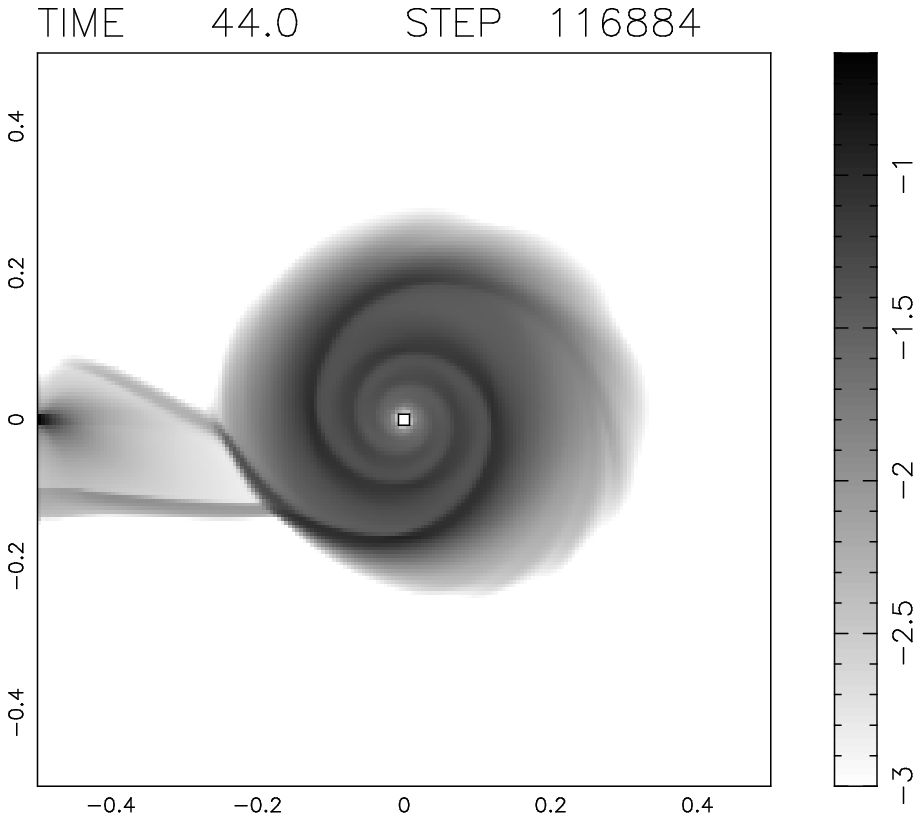, width=50mm}\\
\vspace{-5mm}\hspace{1.2cm}{\Large$\gamma=1.05$} & 
\vspace{-5mm}\hspace{1cm}{\Large$\gamma=1.01$} \\
\end{tabular*}
\end{center}
\caption{Density distribution in two-dimensional calculation. Logarithmic
density of gas is shown at t=44, corresponding to about 7 revolution periods,
for four cases of $\gamma$, {\it i.e.} 1.2, 1.1, 1.05, 1.01. Bar on the right
side shows the scale range.}
\end{figure*}
 The gas is assumed to flow into through a hole placed at the L1 point.  The
density of the gas in the hole is 1 and the sonic velocity is $0.1a \Omega$,
where $\Omega$ is the angular velocity of rotation.  Specifying the sonic
velocity means specifying the gas temperature.  The sonic velocity used in
this calculation is very large, thus indicating that the gas has a
considerably high temperature. The gas ejected through the hole expands into
a surrounding atmosphere having high temperature, low density and low
pressure, thereby forming so called ``under-expanded jet''.  This results in
the strange form of the inflow from the L1 point, which phenomenon turns out,
however, not to be a serious drawback according to our later study.

What the calculations revealed is that the spiral shocks are generated in
any case and that the pitch angle of the spiral arms has a clear correlation
with the specific heat ratio.  A smaller specific heat ratio thus leads to a
tighter winding-in angle of the resulting shock waves.  This is because that the
smaller specific heat ratio causes lower gas temperature and hence smaller
sonic velocity, whereby the spiral shock waves having been generated in the
peripheral region wind in more tightly while propagating into the inner parts. 
In three-dimensional calculations to be described later herein, the problem
is that such a clear correlation is not seen.

\subsection{Spiral shocks in the universe}
The presence of spiral shocks was discovered by Sawada, Matsuda \&
Hachisu(1986a, b, 1987) in accretion disks.  Spiral structures themselves are,
however, rather common phenomena in astrophysics.  In particular, spiral arms
of galaxies attracted much attention in the 60-70's.  Lin \& Shu (1964)
contended that spiral density waves are formed in galactic disks under
influence of self-gravity.  This is the famous density wave theory of spiral
arms.  Fujimoto (1968) suggested that the spiral arms, actually shining, is a
gas rather than the constituent stars and that spiral
shock waves composed of the gas have been formed. Shu, Milione \& Roberts (1973) discussed
the motion of gas in the {\it spiral gravitational potential} formed by stars
and showed that the spiral gravitational potential with its amplitude
exceeding a specific level causes the gas to form spiral shock waves.

S\o rensen, Matsuda \& Fujimoto (1976) conducted numerical simulations for gas
flow in a barred galaxy and showed that shock waves had been formed both
inside and outside the corotation radius.  The outside shock waves are spiral
and correspond to the spiral arms, while the inside shocks are linear and
correspond to the dark lane where cosmic dust has gathered.  

S\o rensen \& Matsuda (1982) and Matsuda et al. (1987) showed that a barred 
galaxy with a small deviation from
axisymmetry of gravitational potential, i.e. a weak bar, causes the shock
waves inside the corotation radius also to become of spiral form.  They
investigated the correlation between the presence of spirals and that of
Lindblad resonance, and concluded that the Lindblad resonance is essential to
generate spirals. An important fact is that spiral shocks are generated by the 
{\it barred gravitational potential} rather than the {\it spiral
gravitational potential}. This mechanism is essentially the same as that for
the formation of the spiral structure in accretion disks of close binaries
discussed in the present paper. 

Matsuda \& Nelson (1977) suggested that the presence of, if any, a weak bar
structure at the center of our Galaxy would cause spiral shock waves to
generate, whereby gas around the waves loses its angular momentum and energy
and falls towards the center.  They called this mechanism ``vacuum cleaner''.
This mechanism is important in considering gas supply to the central core in
AGN.

Besides the galaxy scale, formation of spiral structure due to tidal force is
seen on various objects.  For instance, various simulations have shown, with
the primordial solar nebula after formation of Jupiter, appearance of spiral
shock waves in the gaseous disk. These spiral shocks remove gases from the
primordial solar nebula eventually.

In essence spiral shock waves are generated in an accretion disk 
by an oval deformation of the
gravitational potential and the Lindblad resonances associating with
it. We stress that spiral shocks are not formed by the collision
of the  stream from L1 point with the disk gas.

\subsection{Does a spiral shock appear on three-dimensional scale?}
As described above, the presence of spiral shocks in two-dimensional
accretion disk has been verified by various simulations.  For
three-dimensional disks, it has been argued that waves once formed
in the periphery of a disk diffract upwards and do not move into the inside
of the disk and that, as a result, spiral shocks cannot appear in the disk
(Lin, Papaloizou \& Savonije 1990a, b, Lubow \& Pringle 1993).
 
As regards numerical simulation, Molteni, Belvedere \& Lanzafame (1991) and Lanzafame, Belvedere \& Molteni (1992) conducted three-dimensional simulations by the SPH method and could
not find any spiral shock.  They further argued that, with the specific heat
ratio appearing in the equation of state, \( \gamma \), being larger than
1.1-1.2, accretion disk itself cannot be formed.  Attention should however be
paid to the fact that they used in their calculation particles in as small a
number as 1159 (\( \gamma = 1.2 \)) or 9899 (\( \gamma = 1.01 \)). Since then
for some time, many simulations have been performed mainly with use of the
SPH method (Hirose, Osaki \& Mineshige 1991, Nagasawa, Matsuda \& Kuwahara 1991, Lanzafame, Belvedere \& Molteni 1993), but none of them obtained spiral shocks.

Only in recent years, Yukawa, Boffin \& Matsuda (1997) showed that shock waves are
obtained by the SPH method when the number of particles is increased so that
the resolution is enhanced.  They obtained spiral shocks with the binary
having a mass ratio of 1 and with a specific heat ratio, \( \gamma \), of 1.2,
although they did not with \( \gamma \) of 1.1 or 1.01.  Thereafter,
Lanzafame \& Belvedere (1997, 1998), Boffin, Haraguchi \& Matsuda (1999)
obtained basically similar results.

With respect to three-dimensional calculation by the finite difference/volume
method, Sawada \& Matsuda (1992) performed the calculation with use of the
TVD method and generalized curvilinear coordinates, and obtained spiral
shocks for \( \gamma = 1.2 \).  They calculated, however, only up to half the
orbital rotation period, and hence it is still doubtful if the generation of
shock waves is an established phenomenon or merely a transient one.  Our
group therefore conducted a series of three-dimensional simulations as
described in the following section (see Makita \& Matsuda 1999, Matsuda,
Makita \& Boffin 1999).

Bisikalo et al. have performed a series of three-dimensional calculations
similar to ours (1997a, b, 1998a, b and c).  Their study is very much like ours
and therefore particularly worth commenting.  They used the TVD method and
Cartesian coordinates.  The region of calculation was $[-a, 2a] \times [-a, a]
\times [0,a]$, where $a$ is the separation, and the region was divided with a
non-uniform lattice of $78 \times 60 \times 35$ or $84 \times 65 \times 33$. 
(In our calculation given later, the region is $[-a, a] \times [-0.5a, 0.5a]
\times [0, 0.5a]$, which is divided into $200 \times 100 \times 50$.)  They
used the equation of state of a perfect gas and calculated with the specific
heat ratio \( \gamma = 1.01, 1.2 \).  They calculated over a sufficiently
long period of 12-20 orbital periods.  One of the main points concluded by
them is the absence of ``hot spot'', i.e. high-temperature region generated on
collision of the gas inflowing from the L1 point with the accretion disk.  As
described later, our results also support this conclusion.  They also
conclude that no accretion disk forms with  \( \gamma = 1.2 \), while our
results show the formation.  In addition, they mention, as the cause for the
generation of spiral shocks, rather than the tidal force of the companion,
the interaction between the L1 flow and an expanded atmosphere, to which we
do not agree.
\nopagebreak 
\section{Method of calculation}
\subsection{Basic assumptions}
We consider a mass-accreting star (main star) with mass $M1$ and a
mass-losing star (companion) with mass $M2$. The mass ratio is limited to $q
= M2/M1 = 1$ only in the present work.  The companion is assumed to have
filled the critical Roche lobe.  The motion of the gas flow having 
blown out from the
surface of the companion is calculated by solving the time-dependent Euler
equations.

The basic equations governing the motion of the gas are the Euler equations
with no viscous term.  We consider only numerical viscosity which is
incorporated into calculation from the scheme employed, and consider neither
molecular viscosity nor  \( \alpha \) viscosity.  We further assume that the
equation of state of the gas is expressed by that of a perfect gas and take
the specific heat ratios 1.01 and 1.2.  With accretion disks, radiative
cooling plays an important role.  Judging from the present-day computer power,
it is, however, very difficult to incorporate the effect of the radiative
cooling into three-dimensional calculation.  We therefore try to simulate the
gas cooling to some extent, by using a small specific heat ratio $\gamma$.

The calculation method used is the Simplified Flux Vector Splitting (SFS)
method (Jyounouchi et al. 1993, Shima \& Jyounouchi 1994, 1997).  A
MUSCL-type approach is used, with the calculation accuracy being of 2nd order
for both time and space.

With the centers of the main star and the companion being at $(0,0,0)$ and $
(-1,0,0)$, respectively, the region of calculation is $[-1.5,0.5] \times
[-0.5,0.5] \times [0,0.5]$, where the lengths are scaled by the separation
$a$.  The region is divided by grid points of $201 \times 101 \times 51$. 
The main star is represented by a hole of $3 \times 3 \times 2$, while the 
companion by a surface along the critical Roche lobe.
 
\subsection{Initial conditions and boundary conditions}
\subsubsection{Initial condition}
The entire space except the companion is filled with a gas having a density 
\( 10^{-7} \) and a sonic velocity of $c_0=0.02$.  Here the density is, as
described later, is normalized by that of the gas on the surface of the
companion.  The sonic velocity value is sufficiently smaller than that in Makita, Miyawaki \& Matsuda (1998), who assumed the density and
the sonic velocity to be \( 10^{-5} \) and \( 10^{1/2} \), respectively.  In
their case the gas initially placed in the region of calculation has a low
density, but a high temperature.  These values are not so important in
two-dimensional calculations, since the initial gas will be removed from the
region of the accretion disk eventually.

In three-dimensional calculations, however, the initial gas will partially
remain in a space above the disk.  In the inner regions where the disk is
thin, the disk gas and the initial gas may mix with each other numerically,
thereby increasing the disk temperature artificially.  As a result, inner spiral
arms may wind-in more loosely than in the actual case.  In
order to prevent this, we assume the density and temperature of the initial
gas to be at sufficiently low levels.  

\subsubsection{Boundary conditions}
The inside of the hole representing the main star is always filled with the
above-described initial gas. The gas having reached the vicinity of the main
star therefore accretes due to the low pressure inside the star. The outer
boundary is also fixed at these values.  At the outer boundary, the
artificial reflection of waves is suppressed to a minimum, so that the
calculation can proceed stably. 

The inside of the companion is assumed to be always filled with a gas having
a density of 1 and a sonic velocity of 0.02. The inside of the companion is
assumed to be free from the gravity.  The pressure difference between
the inside and outside of the companion surface causes the inside gas to flow
out, the mass flux of which is evaluated by solving the corresponding Riemann problem.

\section{Results of calculation}
\subsection{Density distribution and spiral shocks}
Figures 2 show the density distribution on the orbital plane at $t=72$,
for \( \gamma \) =1.01 and 1.2, respectively.  Figures 3 show the
iso-density surfaces for the same cases.  The L1 flow, i.e. a flow coming out
of the L1 point, penetrates into the inside of the accretion disk and does
not slow down on collision with the disk or form a ``hot spot''.  
This is discussed in detail later. 
\begin{figure}
\centerline{
\epsfig{file=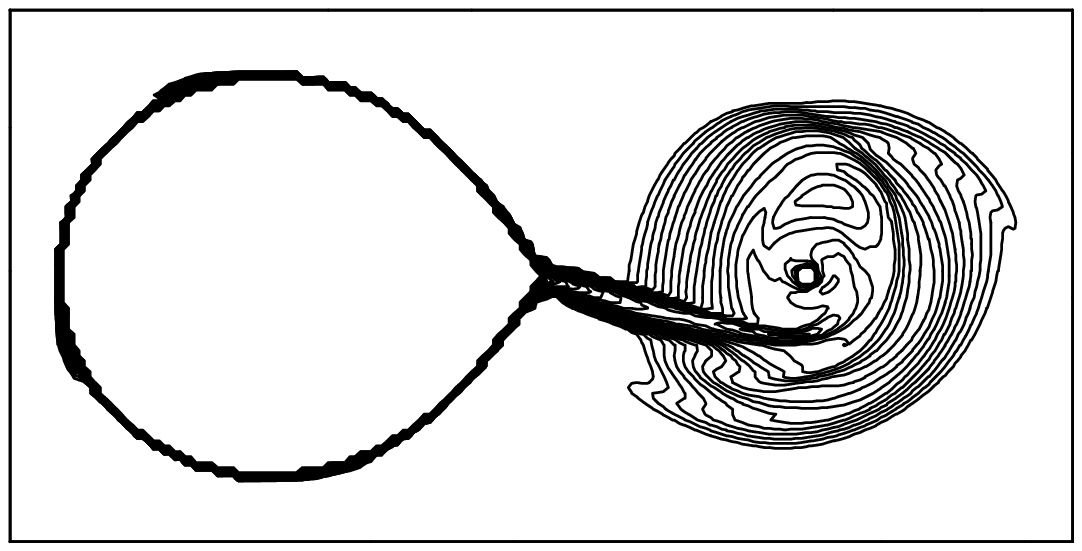,width=55mm} \qquad
\epsfig{file=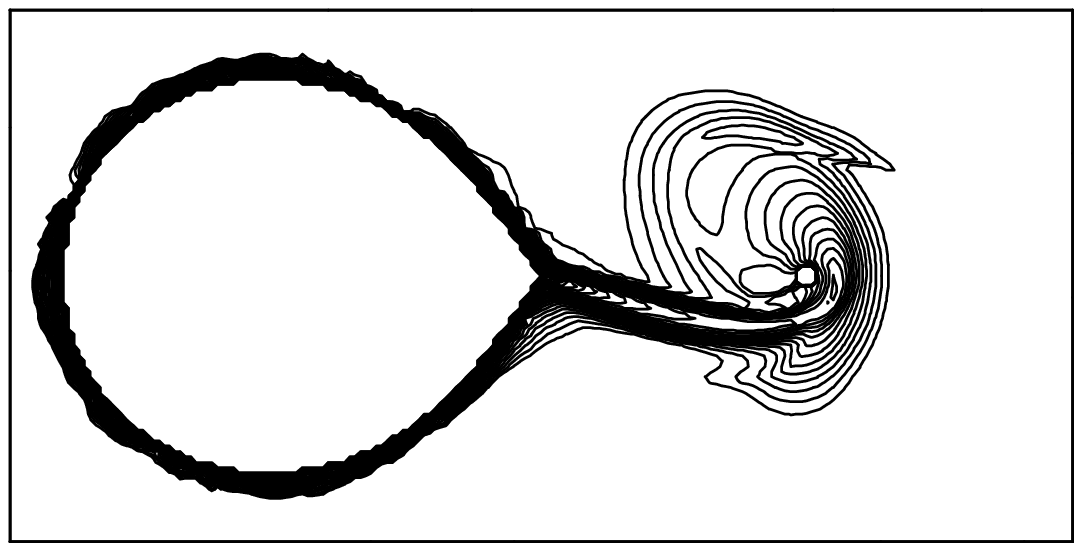,width=55mm}
}
\caption{Density contours in logarithmic scale of the gas on the
rotational plane at $t=72$.  Left: the case for $\gamma=1.01$. Right: the 
case for $\gamma=1.2$.}
\end{figure}
\begin{figure}
\centerline{
\epsfig{file=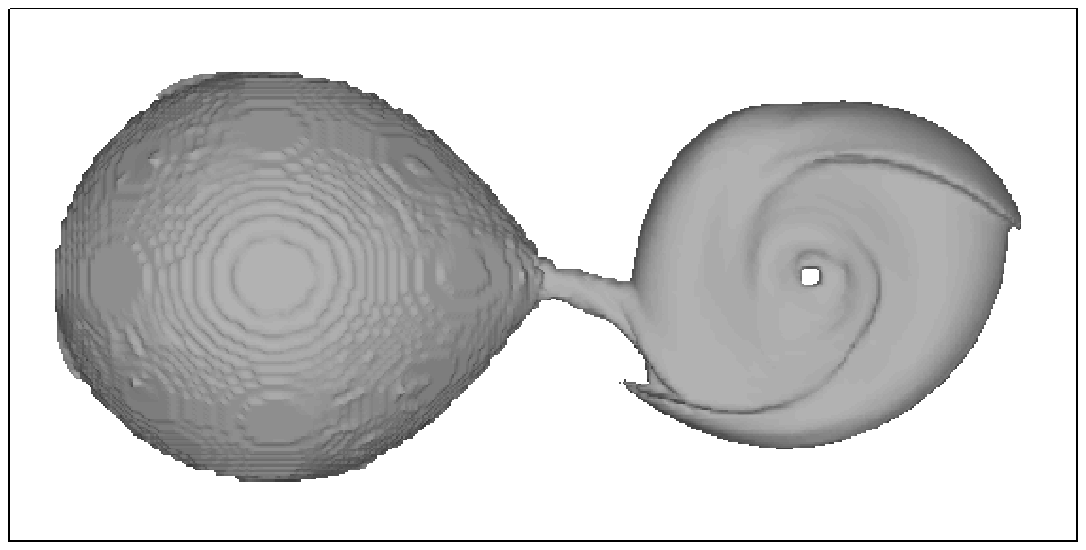,width=55mm} \qquad
\epsfig{file=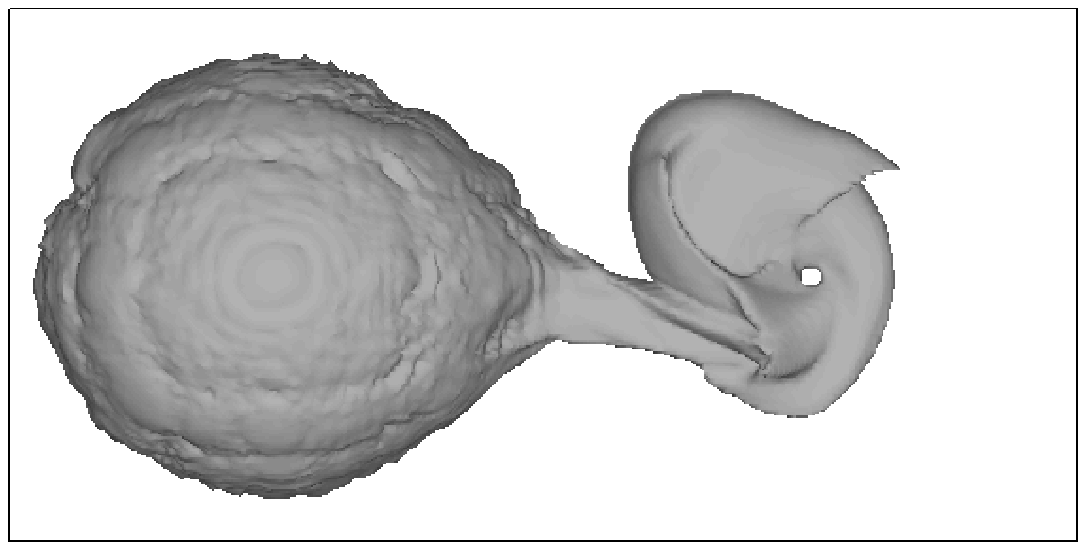,width=55mm}
}
\caption{Iso-density surface at $log \rho=-4.2$. Left: the case for
$\gamma=1.01$. Right: the case for $\gamma=1.2$.
}
\end{figure}
 Another point to be noted is that, while a nearly circular accretion disk
and a pair of spiral shocks are observed for $ \gamma=1.01$, the shape of the
accretion disk is considerably deformed for $\gamma =1.2$.  These results
differ from the preceding results obtained by Makita et al. (1998), who
observed a clearly circular accretion disk and spiral shocks for both
$\gamma=1.01, 1.2$.  Our calculations differ from those of Makita et al. in
the size of calculation region, the mechanism of L1 flow formation, and the
density and temperature of initial gas.  Of these, the former two may not
influence much, and the differences in the conditions of initial gas may have
caused the difference of the results.  Bisikalo et al. (1998a, b, c) argue 
that, for $\gamma=1.2$, a considerably large part of the gas flowing in 
through the L1 point will flow out of the region, so that no accretion disk 
can be formed.
Our results stand in-between those of Makita et al. and Bisikalo et al.  The
question of knowing who is right remains to be answered.

\subsection{Penetration of L1 flow into accretion disk}
It is difficult to visualize a velocity field, which is a vector field.  We
use the Line Integral Convolution (LIC) method to visualize the velocity
field.  For LIC, see Cabral \& Leedom (1993). The details of the 
visualization will be given in a separate paper (Nagae, Fujiwara, Makita, 
Hayashi \& Matsuda, in preparation). 
Figure 4a shows the stream lines on the whole rotational plane.
Figure 4b shows the iso-density surface of $\log \rho =-4.0$ with stream lines
on it. The stream lines on the rotational plane are also shown.
In Figs. 4c and d a portion of
the space between the L1 point and the main star has been expanded, so that
 one can easily see how the L1 flow penetrates into the accretion disk.  As
 seen from Figures 4, the L1 flow coming smoothly through the L1 point
 does, without slowing down on encounter with the disk, penetrate into the
 inside of the disk.
\begin{figure*}
\begin{center}
\epsfig{file=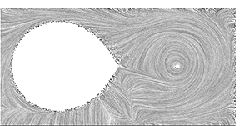,width=9cm}~a. \\
\vspace{5mm}
\epsfig{file=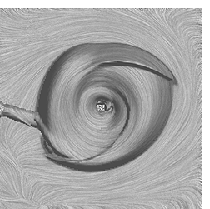,width=9cm}~b. \\
\vspace{5mm}
\epsfig{file=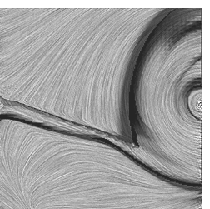,width=5cm}~c. \qquad
\epsfig{file=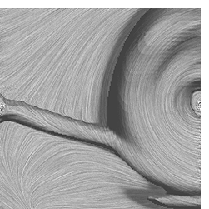,width=5cm}~d.
\end{center}
\caption{Iso-density surface and flow lines on it for the case of
$\gamma=1.01$ at $t=72$. a) Flow lines on the whole rotational plane. 
b) Flow lines on the iso-density surface of $\log \rho=-4.0$ about the mass-accreting star. c) Blow up of L1 stream with $\log \rho=-2.8$, d) Same as c except $\log \rho=-3.3$.
}
\end{figure*}
Figure 4a shows that the rotational flow inside the disk changes, 
on the orbital plane,
its direction rapidly on collision with the L1 flow. Figures 4b, c, d 
show that the
gas placed a little above the orbital plane, along the z-axis, gets over the
L1 flow on collision therewith.  That is, the L1 flow looks like a bar
inserted into the accretion disk, and the disk flow collides with the L1 flow
to form a bow shock on the upstream side (upper side in the Figure).  Figure
4b shows that the iso-density surface swells along a spiral arm. 
Figure 4b shows an iso-density surface with the lowest density, where one can
see that the L1 flow penetrates the accretion disk, like a spear.

Bath et al. (1983) studied the penetration of the L1 flow into the accretion
disk.  Whether or not the L1 flow penetrates into the disk depends on the
relative magnitude of density between the L1 flow and the disk.  An L1 flow
having a larger density will penetrate, while one having a smaller density
will not. Chochol et al. (1984) argued that such a penetration of the L1 stream
was observed in a symbiotic star CI Cyg.

\section{Comparison with observations}
Eleven years after the numerical discovery of spiral shocks by Sawada et al. (1986), Steeghs, Harlaftis \& Horne (1997), using a method called ``Doppler
Tomography'' for observation, discovered the presence of spiral structure on
an accretion disk of IP Pegasi for the first time.  Since then, spiral
structures have been found successively in other accretion disks : SS Cyg (Steeghs et al. 1996), V347 Pup (Still, Buckley \& Garlick 1998), EX Dra (Spruit in preparation).

The Doppler tomography method comprises at first observing the time history
of the emission line spectrum of hydrogen or helium for 1 orbital
period, on a binary system close to edge-on.  The method then makes a
distribution map of the emission lines on the velocity space $(V_x, V_y)$ by
an ingenious technique, the maximum entropy technique. The obtained map is called Doppler map and basically
corresponds to a hodograph in hydrodynamics.  Presence of a spiral distribution
in a Doppler map corresponds to a spiral distribution of emission lines in
the physical (configuration) space.  The shape of the physical-space
brightness distribution, however, cannot be derived from the corresponding
Doppler map. 

On the other hand, numerical simulations can give the physical-space
distribution of a physical quantity, from which a Doppler map can  be
prepared.  Obtaining a distribution of emission lines of an element in the
real space needs both the knowledge of temperature and complex radiative
transfer calculations, and is not easy.  We therefore prepare a Doppler chart,
which maps the density distribution, not the emission line distribution, on
the velocity space.  This procedure may be deemed valid from expectation that
a high-density region be of high brightness.

Figures 5a, b are Doppler maps, prepared in the above-described manner, for \(
\gamma \) =1.01 and 1.2, and show the presence of widely opened spiral
structures.  Matsuda, Makita, Yukawa \& Boffin (1999) and Makita, Yukawa, Matsuda \& Boffin (1999) performed two-dimensional SPH calculations and prepared Doppler
maps. The maps thus prepared agree well with those based on our two-dimensional finite-difference calculations, and also very well with the observations.  

This agreement is remarkable, since the results of these two-dimensional
calculations show that, in particular for \( \gamma \) = 1.2, the obtained
Mach number is as small as less than 10, which means that the disk has a
considerably high temperature.  On the other hand, observations expect much
lower temperatures and a Mach number of 20-30 for the accretion disks of
cataclysmic variables.  For such a high Mach number case, numerical
simulation would give spiral shocks with tightly winding, so that no Doppler
map agreeing with observations can be prepared (Godon, Livio \& Lubow 1998).

Steeghs et al. (1997) observed spiral structures in the outburst phases of a
cataclysmic variable, but not in the quiescent phases.  This suggests that
only with the disk having a high temperature, spiral shocks winding to such
modest angle as to be observable can be formed.
\begin{figure*}
\centerline{
\epsfig{file=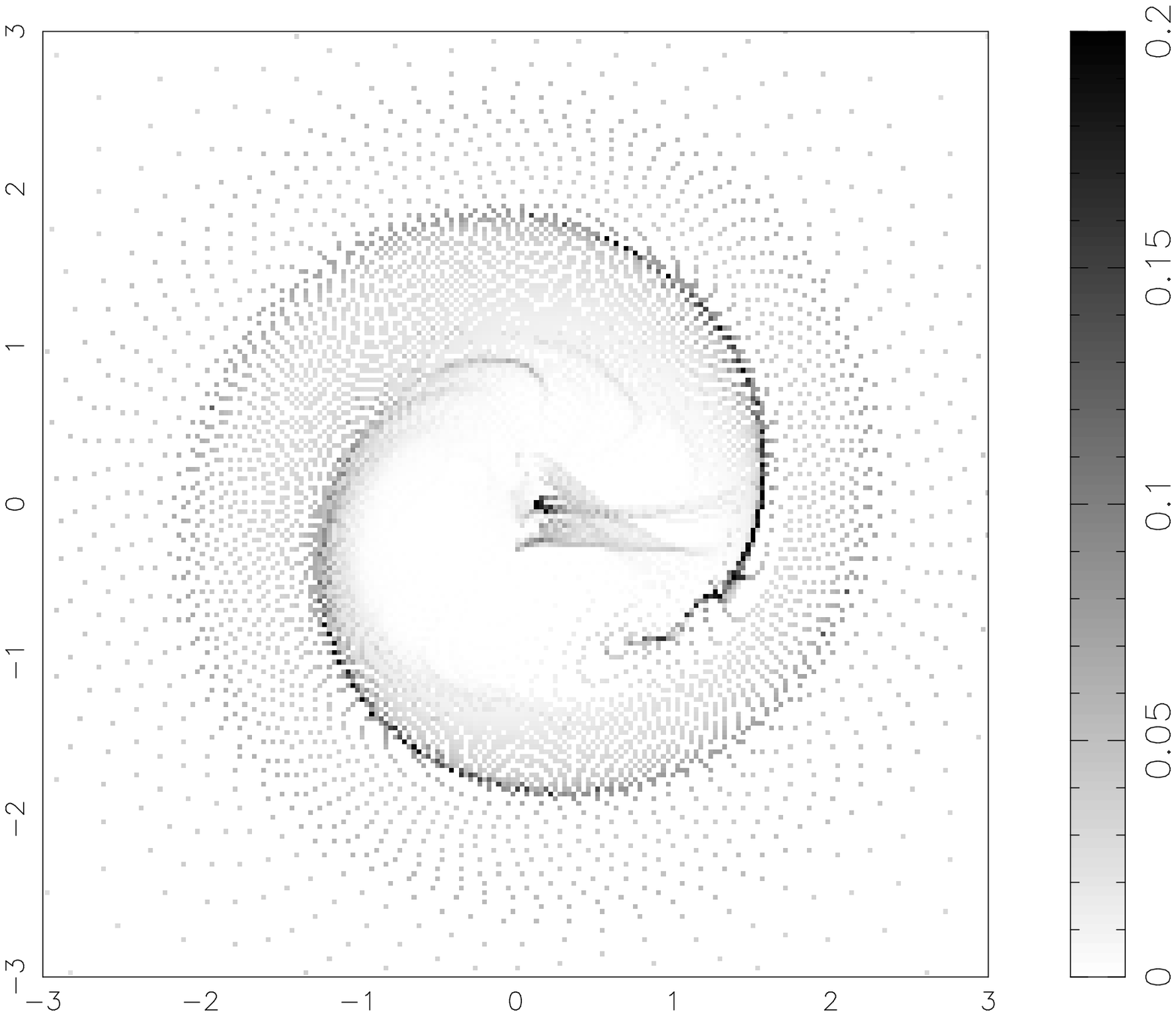, width=48mm}~a. \qquad
\epsfig{file=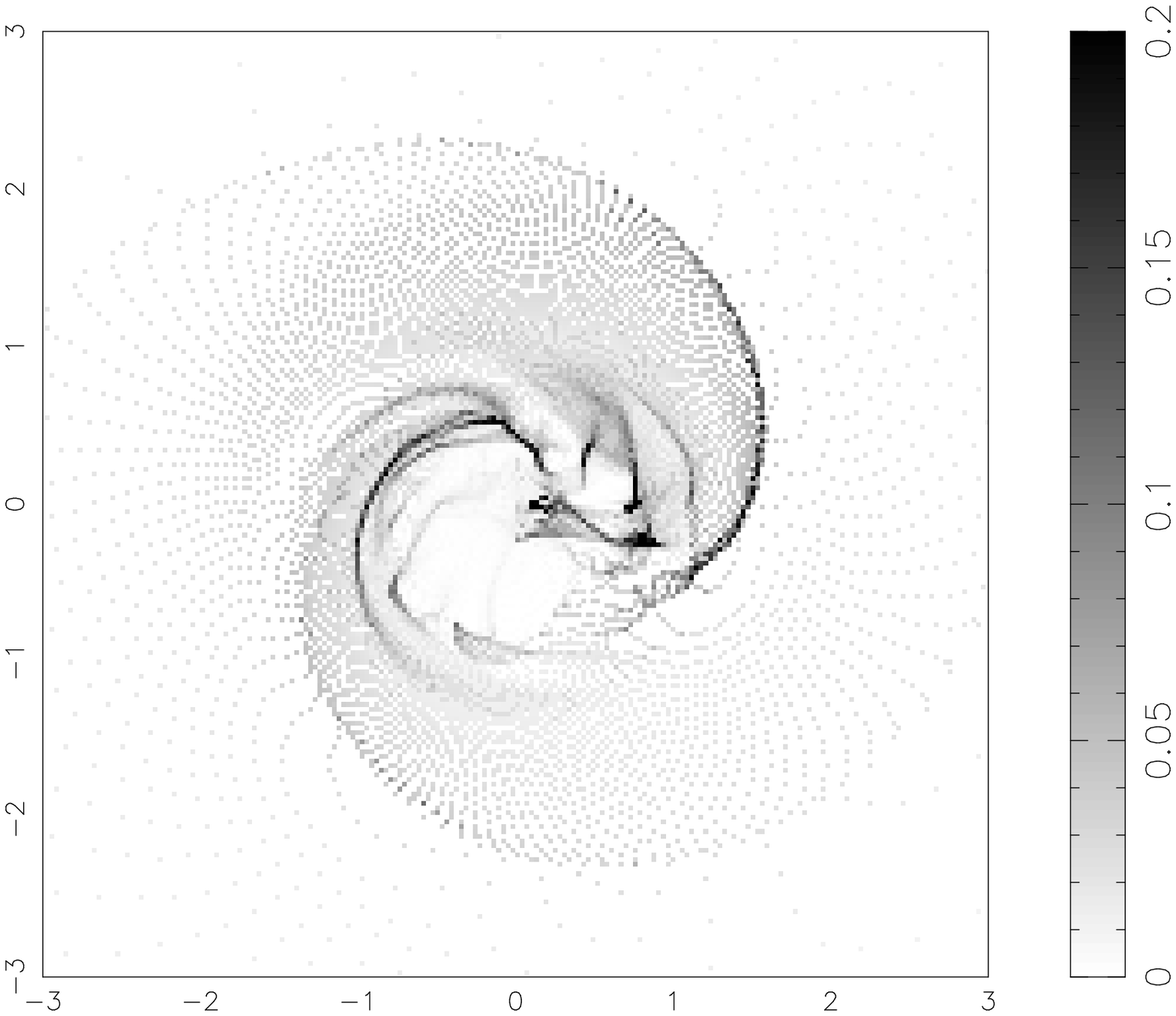, width=48mm}~b.
}
\caption{Doppler map prepared based on two-dimensional calculation; a): 
\( \gamma \) = 1.01 b):  \( \gamma \) = 1.2
}
\end{figure*}
It is hard to prepare a correct, meaningful Doppler map based on the results
of three-dimensional calculations, since the distribution observed is that of
the emission component at an optical depth of 1, i.e. on the photosphere of
the accretion disk.  The preparation thus requires complex calculations of
radiative transfer, which is beyond the scope of the present paper.  One
point of interest here is that the three-dimensional calculations with a
small \( \gamma \) = 1.01 have given moderately loosely winding spiral shocks,
which means that even with accretion disks with comparatively low temperature
the winding-in shape tends to agree with observations.

\section{Summary}
\begin{enumerate}
\item Spiral shock waves on accretion discs in close binary systems were
found numericaly by Sawada et al. (1986a, b, 1987) in their two-dimensional
hydrodynamic finite difference simulation.
\item Spiral shock waves are generated by an oval deformation of the
gravitational potential and the Lindblad resonances associating with
it. Spiral shocks are seen in galactic disks and the primordial solar
nebula as well as accretion disks.
\item Three-dimensional simulations also exhibit the presence of spiral
shocks despite of some counterarguments.
\item The stream from L1 point penetrates into the accretion disk rather
than being blocked by it. Therefore, so-called hot spot is not formed
in our present case,  which fact agrees with the results by Bisikalo
et al. (1997a, b, 1998a, b, c).
\item Spiral structure was found in the accretion disk of a dwarf nova
IP Pegasi by Steeghs et al. (1997) using Doppler tomography technique.
\item Theoretical Doppler maps are made based on two-dimensional simulations,
and they agree well with observed ones despite of high temperature of gas
in the simulation.  
\end{enumerate}

\begin{acknowledgements}
T.M. has been supported by the Grant-in-Aid for Scientific Research
of Ministry of Education, Science and Culture in Japan (11134206)
and (10640231) of JSPS. Calculations were performed on NEC SX-4
at the Data Processing Center of Kobe University and also by
Fujitsu VPP300/16R and VX/4R at the Astronomical Data Analysis Center
of the National Astronomical Observatory, Japan.
\end{acknowledgements}

\end{article}
\end{document}